\documentclass[aps,pre,amsmath,amsfonts,amssymb,floatfix,superscriptaddress,nofootinbib,twocolumn]{revtex4-1}
\usepackage{dsfont}
\usepackage[latin1]{inputenc}
\usepackage{subfigure}
\usepackage{float}
\usepackage{amsmath}
\usepackage[english]{babel}
\usepackage{amsfonts}
\usepackage{amssymb}
\usepackage{fancyhdr}
\usepackage{esdiff}
\usepackage{comment}
\usepackage{textcomp}
\usepackage{cancel}
\usepackage{graphicx}
\usepackage{epsfig}
\usepackage{epstopdf}
\usepackage[T1]{fontenc}
\usepackage{lmodern}
\usepackage{dsfont}
\usepackage{multirow}
\usepackage{tikz} 
\usepackage{stackengine}
\usepackage[font=small]{caption}
\usepackage{epsfig}
\usepackage{epstopdf}
\usepackage{ctable}

\newcommand*{\bd}{\boldsymbol}

\begin{document}

\title{Pattern invariance for reaction-diffusion systems on complex networks}

\author{Giulia Cencetti}
\email[Correspondence to ]{giulia.cencetti@unifi.it}
\affiliation{Universit\`{a} degli Studi di Firenze, Dipartimento di Ingegneria dell'Informazione, Florence, Italy}
\affiliation{Universit\`{a} degli Studi di Firenze, Dipartimento di Fisica e Astronomia and CSDC, Florence, Italy}
\affiliation{INFN Sezione di Firenze, Italia}

\author{Pau Clusella} 
\affiliation{Institute for Complex Systems and Mathematical Biology, SUPA, University of Aberdeen, Aberdeen, UK}
\affiliation{Universit\`{a} degli Studi di Firenze, Dipartimento di Fisica e Astronomia and CSDC, Florence, Italy}

\author{Duccio Fanelli}
\affiliation{Universit\`{a} degli Studi di Firenze, Dipartimento di Fisica e Astronomia and CSDC, Florence, Italy}
\affiliation{INFN Sezione di Firenze, Italia}

\begin{abstract}
        Given a reaction-diffusion system interacting via a complex network, we propose two different  techniques to modify the network topology while preserving its dynamical behaviour. In the region of parameters where the homogeneous solution gets spontaneously destabilized, perturbations grow along the unstable directions made available across the networks of connections, yielding irregular spatio-temporal patterns. We exploit the spectral properties of the Laplacian operator associated to the graph in order to modify  its topology, while preserving the unstable manifold of the underlying equilibrium. The new network is {\it isodynamic} to the former, meaning that it reproduces the dynamical response (pattern) to a perturbation, as displayed by the original system. The first method acts  directly on the eigenmodes, thus resulting in a general redistribution of link weights which, in some cases, can completely change the structure of the original network. The second method uses localization properties of the eigenvectors to identify and randomize a subnetwork that is mostly embedded  only into the stable manifold. We test both techniques on different network topologies using the Ginzburg-Landau system as a reference model. Whereas the correlation between patterns on isodynamic networks generated via the first recipe is larger, the second method allows for a finer control at the level of single nodes. This work opens up a new perspective on the multiple possibilities for identifying the family of discrete supports that instigate equivalent dynamical responses on a multispecies reaction-diffusion system. 
\end{abstract}

\maketitle
%

Networks of interacting elements provide a formal representation to model the structure of many complex systems~\cite{Newman10,LatoraNicosiaRusso17book}, such as social interactions~\cite{Celli_etal10}, transportation models~\cite{CardilloZaninGomezRomanceBoccaletti13}, ecological systems~\cite{Montoya_etal06}, and neuronal functions\cite{Bassett_etal11}.
The dynamical behavior of such systems cannot be explained as a simple superposition of the dynamics of the single units:
one also needs to account for how the different elements interact (type of coupling) and
which connectivity structure links the elements (network topology).
Dynamical systems on complex networks have been widely studied \cite{BarratBarthelemyVespignani08}. However, a general theory to fully resolve the subtle interplay between network structure and ensuing dynamics is still lacking, notwithstanding notable efforts which combine  fundamental \cite{ChengShen10,DeDomenico17,AsllaniCarlettiDiPattiFanelliPiazza18} and applied expertise \cite{Lim_etal14,Clopath_etal17}.

In the particular case of reaction-diffusion systems with identical single node dynamics, the coupling between the units always admits a homogeneous state where all the nodes are synchronized. If such equilibrium proves unstable, any arbitrary perturbation
grows through the unstable directions giving rise to irregular spatio-temporal patterns
\cite{Turing52,OthmerScriven71,Pismen06,AsllaniChallengerPavoneSacconiFanelli14}.
This irregular behaviour represents the \textit{functional} response to a given input which can be traced back to the \textit{structural} characteristics of the underlying network~\cite{Brechtel_etal18}.

Whereas, in general, a random change on the network topology would lead to a modification of its dynamical response, it is also presumable that a specific outcome is not unique from a particular network, so that similar patterns might arise from different \emph{isodynamic} topologies.
The identification of different compatible structures that give rise to the same dynamical behavior represents an important leap forward in the study of complex networks dynamics.
For instance, it paves new roads to devise network reconstruction protocols, in cases where more than one network can correspond to the same dynamical output. Upon analyzing the common topological properties that contain the dynamical information of the system, one might overlook the inaccessible details of the topology.

 In this work we propose two different methods to, given a specific network, generate a second network isodynamic to the first one, so that they share the same dynamical response.
A method to generate isodynamic networks for the heterogeneous Kuramoto model was already introduced in~\cite{ArolaDiazGuileraArenas18}.
Here we assume instead a generic class of systems characterized by identical single-node dynamics and nonlinear diffusive coupling. The system admits a homogeneous equilibrium solution, whose stability reflects the underlying network topology as stored in the spectrum of the associated Laplacian operator~\cite{Nakao2014, Cencetti2017}. 
 We exploit the idea that, in the vicinity of an unstable (homogeneous) equilibrium, only a limited subportion of the overall network architecture proves significant for the emergence of a specific dynamical behavior~\cite{Gottwald15,HancockGottwald18,AndreottiRemondiniBazzani18}. 

The first method is exact, since it directly acts on the subspace generated by the Laplacian eigenvectors characterizing the stability of the equilibrium state.
On the other hand, the second technique relies on a Monte-Carlo algorithm that allows for a neater control in terms of network topology.
In brief, the goal of this work is to prove that it is possible to provide two (or more) structurally different complex networks
so that inserting the same input on them can lead to the same output.

\section{Dispersion relation and pattern formation}\label{dispersionrelation}
Let us consider a generic system composed of $N$ identical entities linked through a complex network.
At time $t$ the activity of node $j$ is described by an $m$-dimensional variable $\bd w_j(t)\in\mathbb{R}^m$.
Starting from a specific initial state $\bd w_j(0)$, the dynamics of $\bd w_j$ evolves according to
\begin{equation}\label{eq_system}
\dot{\bd w_j} = \bd{\mathcal{F}}(\bd w_j) + K\sum_{k=1}^NA_{jk}\bd{\mathcal{G}}(\bd w_k-\bd w_j) \qquad j=1,\dots,N
\end{equation}
where $\bd A$ denotes the adjacency matrix of the network, and $\bd{\mathcal{F}}:\mathbb{R}^m\rightarrow\mathbb{R}^m$ and $\bd{\mathcal{G}}:\mathbb{R}^m\rightarrow\mathbb{R}^m$ are generic continuous functions.
 The first term on the right hand side of the equation describes the self-dynamics of each individual node $j$, and it is denoted as \textit{reaction term}.
 The second term instead, defines the interaction of node $j$ with the other nodes of the network:
 the existence of a coupling is set by the adjacency matrix $\bd{A}$, while the interaction shape is established by function $\bd{\mathcal{G}}$.
 Our analysis requires two assumptions: 
  (i) there exists an equilibrium $\bd{w}_j=\bd w^*$ for the uncoupled problem $\dot{\bd{w}_j} = \bd{\mathcal{F}}(\bd w_j)$, and (ii) function $\bd{\mathcal{G}}$ annihilates in zero ($\bd{\mathcal{G}}(\bd 0)=\bd 0$).
The uncoupled equilibrium $\bd w^*$  can be either a stationary fixed point or a limit cycle.
Whatever the case, the second condition ensures that $\bd{w}^*$ becomes a \emph{homogeneous solution} of the system~\eqref{eq_system}.
A wide class of models corresponding to this setup are \emph{reaction-diffusion} systems, where $\bd{\mathcal{G}}$ is a linear function.
This constraints can be also generalized in order to include networks of homogeneous Kuramoto-Daido systems with arbitrary coupling functions~\cite{Daido-93,Daido-93a,Daido-96}.

At the homogeneous equilibrium $\bd w^*$ no information is flowing through the network.
However, if such state proves unstable, a generic small perturbation $\bd{\delta w}$ of $\bd w^*$ can develop to an irregular spatio-temporal pattern.
A linearization of equation \eqref{eq_system} around $\bd w^*$ provides 
 the time evolution of the perturbation $\bd{\delta w}_j\in \mathbb{R}^{Nm}$ as a system of $Nm$ linear ordinary differential equations,
\begin{equation}\label{eq:jacobian}
\bd{\dot{\delta w}} = \bd{J}(\bd w^*)\bd{\delta w},
\end{equation}
where $\bd{J}\in \mathbb{R}^{Nm}\times \mathbb{R}^{Nm}$ is the Jacobian matrix of the system. 
Therefore, the stability analysis of $\bd w^*$ turns into studying the high-dimensional operator $\bd{J}(\bd w^*)$.
Nevertheless, it is possible to link the diagonalization of the Jacobian to that of a simpler operator.
In the linearized regime, the flow of the quantity $\bd{\delta w}_j$ to another node $i$ of the underlying network
is described by $\Delta_{ij}=A_{ij}-k_i^{(in)}\delta_{ij}$, where $k_i^{(in)}$ is the in-degree of node $i$ and $\delta_{ij}$ the Kronecker delta.
The resulting $N\times N$ matrix $\bd{\Delta}$ is known as \emph{Laplacian} of the network. The Laplacian matrix always has a zero eigenvalue associated to a uniform eigenvector $\bd{\phi}^{(0)}=(\phi_1^{(0)},\dots,\phi_N^{(0)}) \propto(1,\dots,1)$, and all remaining eigenvalues are negative.

By expressing $\bd{\delta w}_j$ in the basis of the Laplacian eigenvectors $\bd{\Phi}=(\bd \phi^{(\alpha)})$,
 and making use of the corresponding eigenvalues $\bd{\Lambda}=(\Lambda^{(\alpha)})$
 it is possible to decouple the $Nm$ equations from system (\ref{eq:jacobian}), thus reducing the problem to $N$ uncoupled $m$-dimensional systems indexed by $\alpha$ (see Appendix). Each of the reduced systems is described by the reduced Jacobian $\bd J_{\alpha} \equiv \bd{\partial_w\mathcal{F}}(\bd w^*) + \bd{\partial_w\mathcal{G}}(\bd 0)K\Lambda^{(\alpha)}$. If $\bd J_{\alpha}$ is time independent, as when $\bd w^*$ is a fixed point, the stability analysis is simply assessed by its $m$ eigenvalues $\bd{\lambda}^{(\alpha)}=(\lambda^{(\alpha)}_k)$.
 Therefore, the stability of $\bd{w}^*$ in the full $Nm$ dimensional problem is ultimately controlled by the relation
 $\lambda(\alpha):=\smash{\displaystyle\max_{k}}(\lambda^{(\alpha)}_k)$.
 If instead the Jacobian has a periodic dependence on time, one needs to obtain
 the $m$ \textit{Floquet exponents} $\bd \mu^{(\alpha)}=(\mu^{(\alpha)}_k)$ of the system~\cite{Grimshaw17,ChallengerBurioniFanelli15,LucasFanelliCarletti18}, which represent the analogue of $\bd \lambda^{(\alpha)}$ for a time-dependent Jacobian (see Appendices).
 In this case, the exponents controlling the stability of $\bd w^*$ are given by $\lambda(\alpha)$ := $\smash{\displaystyle\max_{k}}$ $(\mu^{(\alpha)}_k)$
In both instances, $\lambda$ provides the stability of $\bd w^*$ as a function of the Laplacian eigenvalues. Such relation takes the name of \textit{dispersion relation}.  
If the dispersion relation of a subportion $N_c$ of the total $N$ eigenvalues $\Lambda^{(\alpha)}$ is positive, then any perturbation would grow through the unstable modes.
The resulting irregular spatio-temporal pattern thus represents the unpredictable \emph{response} of the system to a specific \emph{input signal}.
It is however reasonable to suppose that most of the relevant information is stored into the unstable manifold of the homogeneous solution $\bd w^*$, as we will prove later on.

\section{Topology modification}\label{topologymodification}




In this section we propose two methods to modify a given discrete support of the system while preserving the relevant directions for the emergence of the pattern.

\subsection*{Eigenmode randomization}
The first method for network modification consists of preserving a subportion $n$ of the total $N$ eigenmodes of the original network Laplacian whereas all the others are modified.
Let us suppose, without loss of generality, that the subset of modes to be left invariant are the first $n$. 
We define the diagonal matrix $\bd{\tilde\Lambda}$, such that
\begin{equation*}
\tilde \Lambda^{(\alpha)}=\begin{cases}
\Lambda^{(\alpha)}\quad&\text{if}\quad \alpha\leq n\\
\Lambda^{(\alpha)}+\delta\Lambda^{(\alpha)}\quad&\text{if}\quad \alpha > n.
\end{cases}
\end{equation*} 
The corresponding eigenvectors are modified by performing a change of basis 
\begin{equation*}
\tilde{\mathbf{\Phi}}=
\mathbf{\Phi}
\left(
\begin{array}{ c | c}
\mathds{1}_n & \mathbf{0} \\ \hline
\mathbf{0} & \mathbf{R}_{N-n}
\end{array}
\right)
\end{equation*}
where $ \mathds{1}_n$ is the identity matrix of dimension $n$ and 
$\mathbf{R}_{N-n}\in SO(N-n)$ is a random rotation matrix.
In order to practically obtain such rotation we perform QR decomposition of a random matrix.
Taking care of preserving the uniform mode, which corresponds to an identical perturbation acting independently on each node, the transformation
\begin{equation*}
\tilde{\mathbf{\Delta}}:=\tilde{\mathbf{\Phi}}\tilde{\mathbf{\Lambda}}\tilde{\mathbf{\Phi}}^{-1}
\end{equation*}
defines a new Laplacian matrix.

	\subsection*{Local rewiring}

	The previous method does not provide any control whatsoever on the topological modifications introduced in the new Laplacian.
	For this reason, we propose a secondary route that acts at the level of single nodes.
	For many network structures, the Laplacian eigenvectors are well localized on the network,
	\emph{i.e.,} their coordinates in the original vectorial space mostly involve a small subset of nodes, different for each eigenvector~\cite{Hata2017}.
	Therefore, modifying some eigenvectors means acting mainly on the connections of a specific subnetwork, and vice versa.
	With this method we aim to identify and modify links among nodes that are poorly involved on the $n$-dimensional manifold we want to preserve.

	We rely on a Monte-Carlo algorithm that proceeds as follows. Given the original network, we choose a random non-diagonal entry of the adjacency matrix, $A_{ij}$. If the entry indicates the existence of a link between node $i$ and node $j$, such link is removed, otherwise it is created.
	We then compare the $n$ eigenvalues and eigenvectors of the modified network with those of the original one, and only if they prove similar according to a chosen threshold $\tau$ the change is accepted (see Appendix for further details). The process is repeated selecting new random entries over the modified adjacency matrix until a desired number of links have been changed or the method fails to detect new entries that lead to small error.

\section{Results and discussion}\label{results}
\subsection{The Ginzburg-Landau model}
In order to test the different methods we use the Stuart-Landau oscillator as a reference system for the dynamics of each node.
Such model is the canonical form of the Hopf bifurcation, \emph{i.e.},
it describes a generic limit-cycle at the vicinity of a bifurcation where a fixed point becomes unstable through a pair of complex conjugate eigenvalues crossing the imaginary axis.
Therefore, it has been studied in a wide number of applications where periodic oscillatory behavior is relevant.

We consider an ensemble of Stuart-Landau oscillators occupying the nodes of an undirected network with a diffusive coupling (for a case with a directed network see Appendix). 
The equation governing the dynamics of each node $j$, which takes the name of complex Ginzburg-Landau (CGL) equation \cite{vanHarten91, AransonKramer02, Garcia_etal12}, is
\begin{equation}\label{eq:system}
\dot w_j=w_j-(1+ic_2)|w_j|^2w_j+(1+ic_1)K\sum_{k=1}^N\Delta_{jk}w_k
\end{equation}
where $c_1,c_2,K\in\mathbb{R}$ and $w_j\in\mathbb{C}$.
Such system always accepts a homogeneous time-dependent solution,
corresponding to the limit-cycle of a single decoupled element of the network $w_j(t)=w^*(t):=e^{-ic_2t}$.
We focus our analysis on this globally synchronous state.
 
Even though $w^*$ is time-dependent, the Jacobian $\bd J_{\alpha}$ is constant in time, and following the procedure indicated in  the Appendix, 
we obtain the dispersion relation characterizing the linear stability analysis of the limit cycle solution,
\begin{equation}\label{eq:dispersion}
 \lambda(-K\Lambda^{(\alpha)})=K\Lambda^{(\alpha)}-1+ \sqrt{-c_1^2K^2(\Lambda^{(\alpha)})^2+2c_1c_2K\Lambda^{(\alpha)}+1}.
\end{equation}
Depending on the system parameters, function $\lambda$
might have a positive part, corresponding to the smaller Laplacian eigenvalues in absolute value.
Therefore, a particular network with $N$ nodes might have $N_c<N$ Laplacian eigenvalues associated to unstable modes.
From now on, we assume that the Laplacian eigenvalues are sorted in descending order, $\Lambda^{(0)}=0>\Lambda^{(1)}>\dots>\Lambda^{(N)}$,
so that the $N_c$ unstable modes correspond to the Laplacian eigenvalues with index ranging from 1 to $N_c$.

\subsection*{Eigenmode randomization}
\subsubsection*{Erd\"os-R\`enyi networks}

\begin{figure*}
	\centering
	\includegraphics[width=0.9\textwidth]{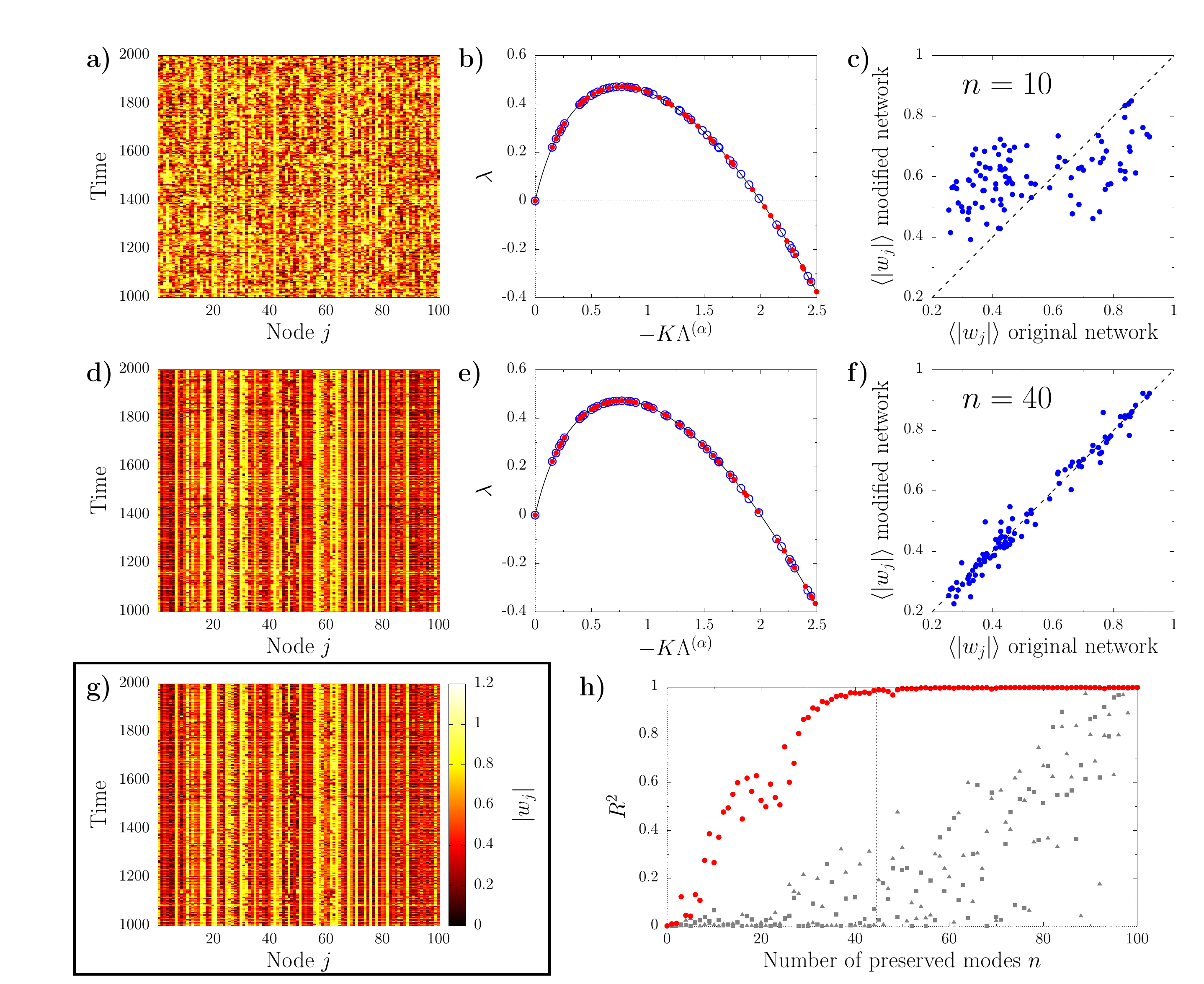}
	\caption{ Results corresponding to an Erd\"os-R\`enyi undirected network with $N=100$
	nodes and average degree $\langle k\rangle=3.5$.
		(a-f) Outcome of the implementation of the eigenmode randomization preserving 10 (a-c) and 40 (d-f) modes.
		(a,d): Time evolution of the modulus for each node of the modified network.
		(b,e): Section of the dispersion relation $\lambda$ showing the unstable eigenvalues for the original (red open circles) and modified network (blue closed circles). A full plot of $\lambda$ for the original network is depicted in the Appendix.
		(c,f): Relation between the time-averaged modulus of each node on the modified network and that of the original network. 
		(g): Time evolution of the modulus of each node of the original network.
		(h): Squared correlation coefficient between the time-average modulus of each node on the modified network and that of the original network for different number of preserved modes.
		Purple circles correspond to networks where the first $n$ Laplacian eigenvalues are preserved. Grey squares and grey triangles are the outcome of two different realizations where the $n$ preserved modes of the modified networks are selected at random.
	}
	\label{fig:panel_ER}
\end{figure*}

We first test the eigenmode randomization method using as original topology an Erd\"os-R\`enyi (ER) network with $N=100$ nodes and average node degree $\langle k\rangle=3.5$. For system parameters $K=1$, $c_1=1$, and $c_2=-3$, the dispersion relation associated to this network
has $N_c=44$ modes corresponding to unstable directions (see blue open circles in Fig.~\ref{fig:panel_ER}).
We integrate the system using as initial condition a randomly generated small perturbation of the homogeneous limit-cycle solution.
After a short transient, the system reaches a dynamical regime characterized by an irregular spatio-temporal pattern depicted in Fig.
~\ref{fig:panel_ER}(g).
Notice that throughout the paper we focus only on each oscillator modulus,
since the frequency of rotation does not seem to contain relevant spatial structure (see Appendix for more details).

We aim to generate a new topology that reproduces such dynamical behavior by leaving invariant a subset of the original network modes.
Taking into account that the Laplacian eigenvalues are sorted in descending order,
we preserve the first $n$ modes and modify the rest using the eigenmode randomization method.
As an illustrative example, in Fig.~\ref{fig:panel_ER}(a) we show the pattern resulting
from a network where only the first $n=10$ modes of the original topology are preserved and all the others have been randomized.
Manifestly, such outcome has little similarity with that of the original network (\emph{cf.} with Fig.~\ref{fig:panel_ER}(g)).
In Fig.~\ref{fig:panel_ER}(b) we plot a zoom on the unstable part of the dispersion relation for the modified network (red dots),
and that of the original topology (blue open circles).
Overall is plausible to explain the disagreement between the dynamical behavior of the two networks from the large difference between the corresponding tangent space of the synchronized solution.
Repeating the procedure with $n=40$ preserved modes instead, leads to the pattern
from Fig.~\ref{fig:panel_ER}(d). In this case,
the behavior of the modified topology resembles much better that of the original network, an observation
which is to be traced to the similarity between the dispersion relations of the respective unstable manifolds (see Fig.~\ref{fig:panel_ER}(e)).

\begin{figure*}
	\centering
	\includegraphics[width=\textwidth]{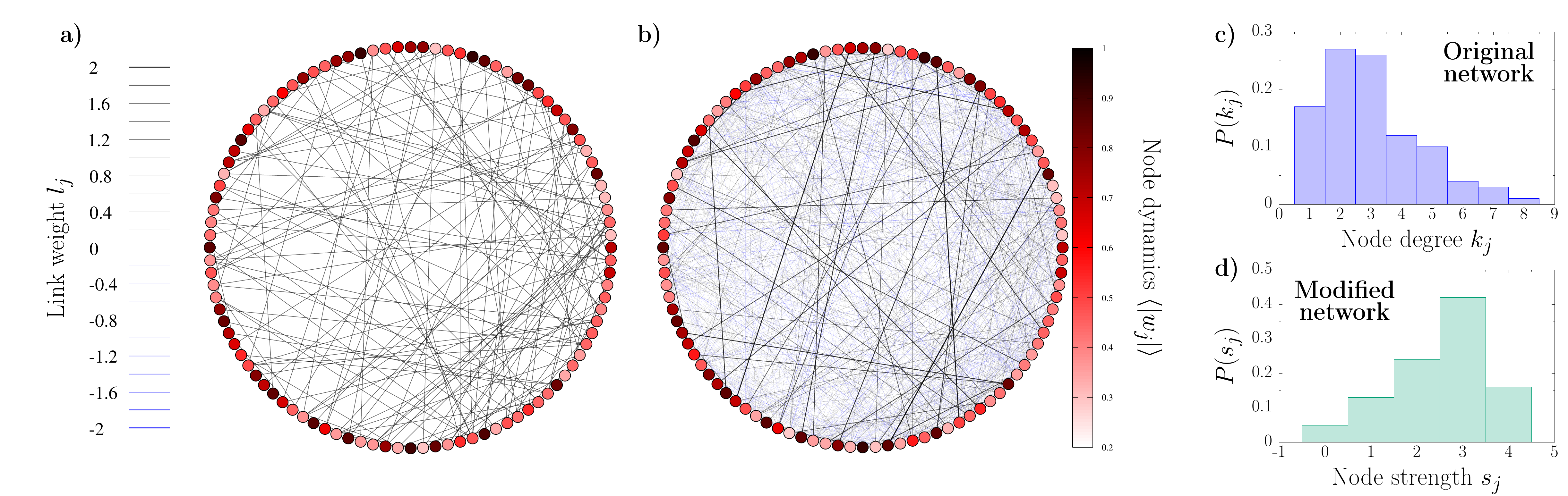}
	\caption{
	(a) Original ER topology. 
	(b) Network resulting from preserving all the unstable modes.
	The color of the nodes correspond to the time average modulus of $W_j$,
	each link thickness is associated to its weight and blue links indicate negative contributions.
	(c) Node degree distribution of the original network. (d) Node strength of the modified network.
	}\label{fig:topologyER}
\end{figure*}

In order to make the comparison between different patterns more transparent,
we compute the time-average modulus of each node, $\langle |W_j|\rangle$ after discarding a transient of 1000 time units.
In Figs.~\ref{fig:panel_ER}(c) and (f) we report the resulting mean node activity of the modified networks versus that of the original one
for $n=10$ and $n=40$ respectively.
An outcome from two identical patterns would lie exactly on the diagonal,
while two completely independent processes would provide a random collection of points.
One can then quantify the similarity between the modified and original patterns in terms of squared Pearson correlation coefficient $R^2$,
which is $0.38$ for $n=10$ and $0.97$ for $n=40$.
In Fig.~\ref{fig:panel_ER}(h) we show
the outcome of repeating this analysis systematically for increasing number of preserved modes $n$.
When the modes to be preserved are selected to be the first $n$, the correlation between the patterns of the modified and original networks increases quickly with $n$, reaching a very good agreement when approximately all the 44 unstable modes are preserved (see red circles).
On the other hand, if the $n$ invariant modes are randomly selected among all the $N$,
even when a large number of directions are maintained, there is no
guarantee that the modified network will respond similarly to the original one (see gray triangles and squares).

The eigenmode randomization technique does not provide, \textit{a priori}, any information about the structure of the resulting network.
In Figs.~\ref{fig:topologyER}(a) and (b) we show the topologies of the original network and a modified version where
all the 44 unstable modes have been preserved.
The first obvious difference is the existence of weighted links, including negative ones (see blue edges), which are absent in the original network. The adjacency matrix loses its sparsity and the network becomes highly connected, although most of the new links are very weak. In order to compare the two networks we focus on the \emph{strength} $s_j$ of each node, defined as
$s_j=\sum_{m=1}^N A_{jm}$, which extends the concept of node degree to weighted topologies.
In Fig.~\ref{fig:topologyER}(c) and (d) we plot the degree distribution of the original network, and the strength distribution of the modified topology, respectively.
Although the distribution of the new network looks different from the original one, it does not present any particular structure, as expected for a random topology.
In fact, the average strength of the new network coincides with the average degree of the initial support.

\subsubsection*{Scale-Free networks}
\begin{figure} [h!]
	\includegraphics[width=0.5\textwidth]{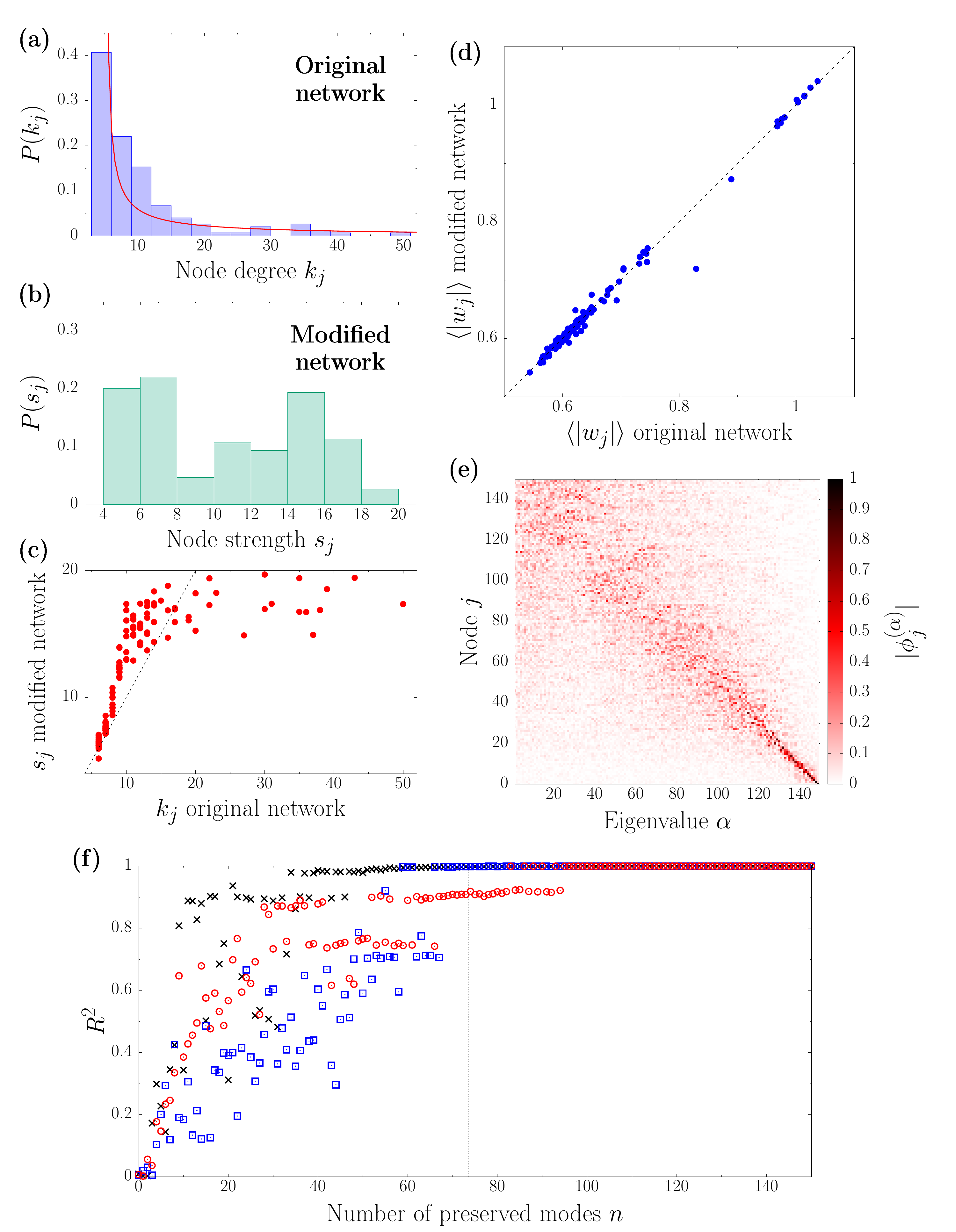}
	\caption{ 
		Outcome of the eigenmode randomization technique using as original network a  scale-free topology with $N=150$ nodes. 
		(a) Degree distribution of the original network (blue boxes). The continuous red curve corresponds to a numerical fit of the data to an exponential shape $P(k)=c k^{-\beta}$, resulting to $c=0.23$ and $\beta=0.85$. 
		(b) Strength distribution on a modified version of the network where the $76$ unstables modes have been preserved.
		(c) Node strength $s_j$ of the modified network with respect to the node degree of the same node $k_j$ in the original network. The black dashed line shows the case $s_j=k_j$ for eye guide.
		(d) Correlation between the dynamical patterns of the original and modified networks, with resulting squared correlation coefficient $R^2=0.99$.
		(e) Eigenvectors of the original Scale-Free network Laplacian.
		The eigenvalues are sorted in descending order according to the corresponding Laplacian eigenvalues, and the nodes 
		have been sorted in descending order with respect to their degree.
		The color represents the absolute value of each eigenvector component.
		(f) Squared correlation coefficient as the number of preserved modes $n$ increases. Red circles, blue squares and black crosses indicate
		the results corresponding of three different initial conditions. The vertical dashed line indicates the threshold between unstable and stable modes.
	}\label{fig:panelBA}
\end{figure}

In order to further investigate the relation between the Laplacian eigenmodes and network topology, we next move to a case where the network has a specific structure.
For this purpose we use Scale-Free networks (SF), characterized by a typical power law degree distribution.
In Fig.~\ref{fig:panelBA}(a) we show an example of such degree distribution for a network composed of 150 nodes built using the Barabasi-Albert~\cite{BarabasiAlbert99} preferential attachment algorithm.
Fixing the parameters of the CGL equation as $K=1$, $c_1=1.2$, and $c_2=-10$, the dispersion relation of this network shows $N_c=76$ unstable modes.
We apply the eigenmode randomization technique to generate a new support taking care to preserve all the unstable directions, but none of the stable ($n=N_c$).
The patterns resulting from inserting the same perturbation to the synchronized solution on both networks are highly alike, as can be
seen from the correlation Fig.~\ref{fig:panelBA}(d).
Differently from the ER case, here the modulus of each node gets stationary after some transient,
so such time-average activity is free from statistical fluctuations.

The topology of the new network is, again, highly connected and involves negative links.
Morevoer, the node strength distribution (see Fig.~\ref{fig:panelBA}(b)) does not preserve the scale-free structure of the original topology.
To identify the nature of this strong changes, in Fig.~\ref{fig:panelBA}(c)  we plot a one to one comparison between the degree of each node of the original support and the corresponding strength on the new topology.
This analysis reveals that nodes with smaller degree on the original setup have a comparable strength after the modification, whereas the hubs become much weaker.

The localization properties of the Laplacian eigenvectors allow us to understand this situation. Inspired by the analysis performed by Hata and Nakao in \cite{Hata2017}, we depict in Fig.~\ref{fig:panelBA}(f) the absolute values of the vector components $|\phi_i^{(\alpha)}|$, where nodes are sorted according to their degree in descending order ($k_1\geq k_2\geq...\geq k_N$), whereas the eigenmode index $\alpha$ follow the usual descending order of the eigenvalues.
 The almost diagonal behavior indicates that the eigenvectors associated to the eigenvalues with smaller absolute value mostly involve the less connected nodes of the network. 
 
This implies that all the networks generated with the eigenmode randomization procedure will mostly differ from the original one for what concerns the nodes characterized by a large degree $k_j$.
For this reason the characteristic shape of the degree distribution is not preserved  in the second network. 
Nevertheless, in dynamical systems with a different shape of the dispersion relation 
it might be possible to maintain the tail of the distribution if the stable modes are to be found among the first Laplacian eigenmodes (see a specific example in the Appendix).

Finally, we repeat the procedure of systematically increasing the number of invariant modes $n$ from 1 to 150 and tracking the resulting correlation coefficient, Fig.~\ref{fig:panelBA}(f).
For each new network we analyze the dynamical patterns resulting from inserting three different randomly generated perturbations. As for the ER case, the correlation between the dynamics of the original and modified topologies increases quickly with $n$, being nearly optimal when all the unstable modes are preserved (see vertical dashed line). Nevertheless, the choice of the initial condition is clearly relevant:
whereas in one case preserving around $40$ modes already provides a good agreement (see black crosses), in other situations one might need to preserve also a considerable number of stable directions to reproduce the original pattern (see red circles).

\subsection*{Local rewiring}

The second method we present offers the advantage of a larger control on the resulting network topology so that, for instance, one can keep the network binary. 
On the other hand, one needs to allow for some variability also on the unstable manifold due to the non perfect localization of the Laplacian eigenmodes on a specific subnetwork.

\begin{figure}
	\centering
	\includegraphics[width=0.47\textwidth]{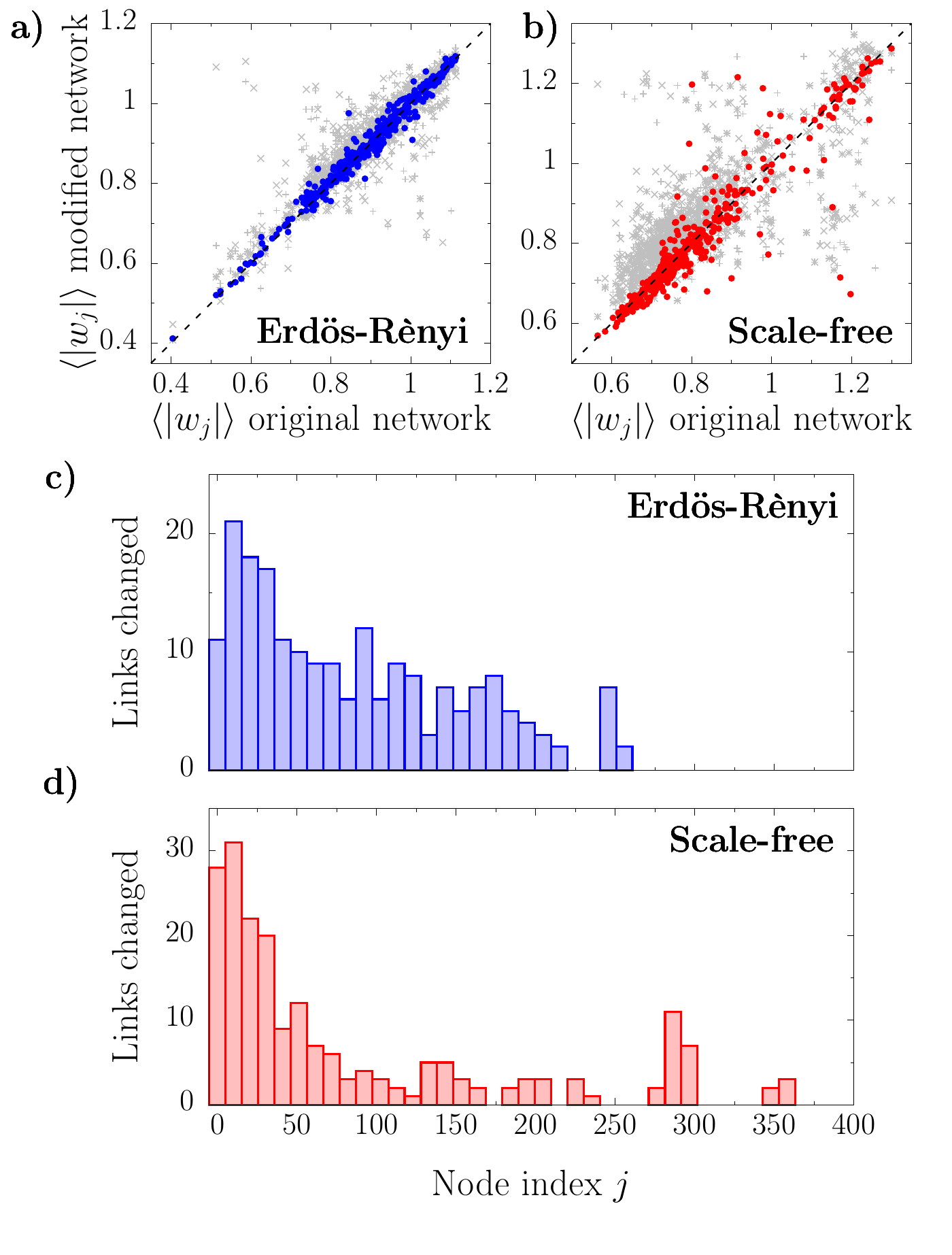}
	\caption{Results of the local rewiring method applied to a ER with average degree $\langle k \rangle$=20 and 64 unstable modes (a,c), and to a SF network with 63 unstable modes (b,d). Both topologies consist of $N=400$ nodes. At each step the tolerance error of the algorithm is $\tau=0.1$. In total, the new networks present 100 modified links with respect to the original topologies. (a,b) Correlation between patterns of the original and modified networks. Colored circles correspond to the results obtained using the local rewiring algorithm. Gray pluses, crosses, and stars correspond to three different networks generated by random rewiring as many links as the network outcoming from the local rewiring procedure. The squared coefficient correlations for the modified ER network is 0.98 (blue circles) whereas the randomly rewired networks provide 0.67, 0.62, and 0.94. The modified SF network has $R^2=$0.86 (red circles), whereas the random networks provide 0.49, 0.50, and 0.67. (c,d) Number of link changed for each node of the network. The nodes are sorted in descending degree order.
	}\label{fig:panel_rewire}
\end{figure}

We first apply the local rewiring technique on a ER network with $N=400$ nodes
and average degree $\langle k\rangle=20$.
For system parameters $K=0.15$, $c_1=1$, and $c_2=-3$, the system displays 64 unstable modes. 
The algorithm stops after 100 positions on the adjacency matrix have been changed.
For comparison purposes, we also created three different networks where 100 entries on the adjacency matrix have been changed totally at random.
Blue circles in Fig.~\ref{fig:panel_rewire}(a) show the correlation between the patterns of the original and the resulting network, whereas the gray symbols correspond to that
of the random rewired graphs.
The dynamical behavior of the network obtained using the local rewiring method provides a better agreement with the original pattern.
We obtain similar results when using SF network generated with the Barab\'asi-Albert algorithm (see Fig.~\ref{fig:panel_rewire}(b)).
Although these results change upon inserting different initial perturbations, the rewiring technique outperform the random modified networks in most of the cases (see Appendix for additional results).

In terms of topology changes, in Figs.~\ref{fig:panel_rewire}(c) and (d) we plot the total number of links that have been modified for each node of the network.
It is clear that most of the changes correspond to nodes with larger degree which, as shown  in the previous section, mostly involve the Laplacian eigenvectors associated to stable directions in the case of the CGL equation. 

\section{Conclusions}\label{conclusions}

In the present work we have devised two specific strategies of network generation so as to mimic the pattern obtained by a former sample network when both are subject to the same non-linear reaction-diffusion system. 
The first method is analytical, and provides isodynamic
networks at the cost of changing important topological features of the original graph.
The second technique instead makes use of a Monte-Carlo algorithm, allowing for a larger control in terms of network topology, but providing less accurate results.
Both methods rely on a preliminary identification of a manifold generated by the Laplacian eigenvectors associated to homogeneous solution instability, which has to be preserved during the modifications. 

This work sheds light on the strong dependence between network topology and dynamical patterns.
The localization properties of the Laplacian eigenvectors unveil the existence of underlying topological features that lead the dynamical response of the network.
In particular, it is possible, in some cases, to identify a subnetwork or a set of nodes that can be acknowledged as practically irrelevant for pattern formation.
It is then clear that distinct networks with differences solely localized on these particular substructures can, under apt conditions, generate the same irregular patterns.
As a remarkable example of this situation, we have shown that the SF networks, often taken as reference model to describe many real systems, can be, in the case of the Ginzburg-Landau system,
replaced by graphs characterized by highly different structures.

Arguing on this line, the two methods proposed here may be placed as novel schemes for network generation, with a purely functional angle:
instead of building the specific network structure brick by brick, paying attention to local characteristics of nodes, the complex graph is  globally established  and constrained to yield an \textit{a priori} chosen dynamical output.\\



As already mentioned, the issue of generating isodynamics networks has been tackled in~\cite{ArolaDiazGuileraArenas18} with reference to the heterogeneous Kuramoto model. In this latter work, the focus was placed on preserving global collective variables, as e.g. reflecting  the displayed level of synchronization. At variance, we here deal with a local definition of "dynamical invariance", which extends to individual nodes. 
In carrying out the analysis, we have however postulated the existence of an underlying homogeneous equilibrium, a working hypothesis which makes the method unsuited to deal with quenched disordered, as in~\cite{ArolaDiazGuileraArenas18}. Extending the proposed technique to include the case of heterogeneous equilibria is an intriguing direction of investigation which is left for future work.

\vspace{1cm}
\noindent \textbf{Author contributions}: G.C. and P.C. performed the calculations and the numerical simulation, G.C., P.C. and D.F. analyzed the results and wrote the manuscript.\\
\textbf{Competing financial interests}: The authors declare no competing interests.

\bibliographystyle{ieeetr}
\bibliography{references_patterninvariance}

\end{document}